\documentclass[A4paper,unsortedaddress,superscriptaddress,prb,twocolumn]{revtex4-1}

\usepackage{hyperref}
\usepackage{epsfig}
\usepackage{graphicx}
\usepackage{subfigure}
\usepackage{latexsym}
\usepackage{color}
\usepackage{fullpage}
\usepackage{dcolumn}
\usepackage{bm}
\usepackage{ulem}
\usepackage{units}

\usepackage{units}

\begin{document}

\title{Local deformations and incommensurability of high quality epitaxial graphene on a weakly interacting transition metal}%

\author{Nils Blanc}
\affiliation{CEA-UJF, INAC, SP2M, 17 rue des Martyrs, 38054 Grenoble Cedex 9 -- France}
\author{Johann Coraux}
\email{johann.coraux@grenoble.cnrs.fr}
\affiliation{Institut N\'{E}EL, CNRS \& Universit\'{e} Joseph Fourier -- BP166 -- F-38042 Grenoble Cedex 9 -- France}
\author{Chi Vo-Van}
\affiliation{Institut N\'{E}EL, CNRS \& Universit\'{e} Joseph Fourier -- BP166 -- F-38042 Grenoble Cedex 9 -- France}
\author{Alpha T. N'Diaye}
\affiliation{II. Physikalisches Institut, Universit\"{a}t zu K\"{o}ln, Z\"{u}lpicher Str. 77, 50937 K\"{o}ln -- Germany}
\author{Olivier Geaymond}
\affiliation{Institut N\'{E}EL, CNRS \& Universit\'{e} Joseph Fourier -- BP166 -- F-38042 Grenoble Cedex 9 -- France}
\author{Gilles Renaud}
\affiliation{CEA-UJF, INAC, SP2M, 17 rue des Martyrs, 38054 Grenoble Cedex 9 -- France}

\date{\today}%

\begin{abstract}

We investigate the fine structure of graphene on iridium, which is a model for graphene weakly interacting with a transition metal substrate. Even the highest quality epitaxial graphene displays tiny imperfections, \textit{i.e.} small biaxial strains, ca. 0.3\%, rotations, ca. 0.5$^\circ$, and shears over distances of ca. 100 nm, and is found incommensurate, as revealed by X-ray diffraction and scanning tunneling microscopy. These structural variations are mostly induced by the increase of the lattice parameter mismatch when cooling down the sample from the graphene preparation temperature to the measurement temperature. Although graphene weakly interacts with iridium, its thermal expansion is found positive, contrary to free-standing graphene. The structure of graphene and its variations are very sensitive to the preparation conditions. All these effects are consistent with initial growth and subsequent pining of graphene at steps. 

\end{abstract}

\maketitle

\section{Introduction}

Graphene preparation at the surface of low-carbon solubility metals like Ir,\cite{Coraux2009} Cu,\cite{Li2009} or Pt\cite{Gao2012} is a surface-confined process which stops once the surface is passivated by a full graphene layer. Therefore it is a straightforward route towards the production of large-area, highly quality graphene. Not only is the carbon solubility low in the aforementioned metals, but also their interaction with graphene. In particular, in the model graphene on Ir(111) system, where graphene was shown to be almost free-standing,\cite{Busse2011} some of the longstanding issues in graphene research could recently be addressed, noteworthy the engineering of graphene's Dirac cone via bandgap opening\cite{Pletikosic2009,Balog2010} or Fermi velocity renormalization,\cite{Rusponi2010} electron confinement in graphene quantum dots,\cite{Phark2011,Hamalainen2011,Subramaniam2012} or the use of graphene as a weakly-perturbating and protective layer for fragile surface phenomena such as the Rashba-split Ir surface state.\cite{Varykhalov2012} The weak interaction of graphene with metals such as Ir, Cu or Pt also has important consequences regarding the structure of graphene. Indeed, graphene domains with different stacking with respect to the substrate tend to form on these metals, presumably because the formation of each of the domain involves similar energetic costs. This was first observed some decades ago,\cite{Land1992} further addressed in details on Ir,\cite{Coraux2008,Loginova2009} and latter on, on Cu,\cite{Wofford2010,Nie2011} and Pt.\cite{Sutter2009a,Merino2011} At the boundary between these domains, dislocations (heptagon-pentagon pairs) are found.\cite{Coraux2008}

Such structural defects hinder the improvement of graphene’s performances, \textit{e.g.} for electronic transport,\cite{Yu2011,Petrone2012} towards state-of-the-art ones obtained for exfoliated, suspended graphene. Certain defects are on the contrary desirable: some point defects are expected to switch on magnetism in graphene,\cite{Yazyev2008} inhomogeneous strain fields were shown to induce electronic gap opening,\cite{Levy2010} and defects are thought a pathway for the intercalation of various species, which recently appeared a powerful route for preparing graphene hybrid systems.\cite{Varykhalov2008,Weser2011,Meng2012,Coraux2012} Producing defect-free single-crystalline graphene or graphene with well-known defects having specific properties on metals such as Cu or Ir is thus of both fundamental and applied interest. Recently, strategies have been developed in order to avoid the formation of grain boundaries, in graphene grown on Ir(111).\cite{vanGastel2009,Hattab2011} Scanning tunneling microscopy (STM), low energy electron diffraction, and low-energy electron microscopy allowed concluding to a single crystalline orientation in graphene at the scale of a centimeter; however, the question of graphene's crystalline perfection can only be answered with higher resolution probes, such as X-ray scattering, which is also an unvaluable tool to gain knowledge about the nature of defects.

We report here the investigation of the structural properties in graphene/Ir(111), by \textit{in situ} high resolution X-ray diffraction, which allowed us to unveil, together with STM, small deviations to a perfect graphene lattice and provide insight into the nature of the epitaxy between graphene and the metal.

The article is organized as follows. After presenting the experimental methods, we turn to the experimental results, and successively address the average lattice parameter in graphene, the distributions of the structural parameters of graphene, and the dependence of these parameters on temperature. The last section of the article is devoted to the discussion of these results.

\section{Methods}

X-ray diffraction with synchrotron light was conducted inside the ultra-high vacuum growth chamber, under grazing incidence to achieve maximum sensitivity to the graphene overlayer; STM was performed in a separate system.

The iridium single crystal polished in a (111) plane, bought from the "Surface Preparation Laboratory", was prepared by repeated cycles of 0.8-1.5~keV Ar$^+$ sputtering at room temperature (RT) and high temperature flash. In order to decrease the crystal mosaic spread and increase the terrace width (of the order of 100~nm after surface preparation), flash temperatures as high as 1770~K were employed. Repeated sample annealing at 1070~K under O$_2$ partial pressures of 10$^{-8}$~mbar, during 2~h, allowed removing carbon incorporated into the sample during high temperature flashes. The sample temperature $T_\mathrm{s}$ was measured with a pyrometer (60~K uncertainties).

\begin{figure}[htpb]
  \begin{center}
  \includegraphics[width=70mm]{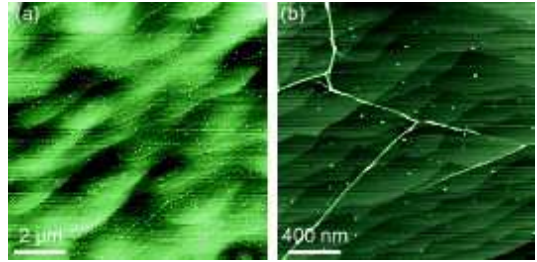}
  \caption{\label{fig:AFM}(Color online) (a,b) Atomic force microscopy images measured in atmospheric conditions on an Ir(111) single-crystal covered with graphene grown under ultra-high vacuum. Note the interconnected network of wrinkles, which reveals the presence of graphene over the whole surface.}
  \end{center}
\end{figure}

Graphene growth was performed by first adsorbing ethylene at RT during 5~min with a 10$^{-7}$~mbar pressure of ethylene in the chamber, then flashing the sample temperature to 1470~K, and finally decreasing the sample temperature to $T_\mathrm{g}$ = 1070-1270~K and letting in a 10$^{-8}$~mbar pressure of ethylene for 10~min.\cite{vanGastel2009} This is known to yield a single layer graphene sheet with almost 100\% coverage, as shown on the atomic force microscopy images in Fig.~\ref{fig:AFM}, free of defects such as grain boundaries, with a single crystallographic orientation, at least to the accuracy of STM and LEED. Compared to growth above 1270~K without the first step (RT absorption and flash anneal at 1470~K), it also reduces the lattice mismatch between graphene and its substrate at RT after growth, and thus limits the density of wrinkles which form upon cooling. Platinum was deposited at RT on graphene/Ir(111) using electron-beam evaporation of a high purity Pt rod under ultra-high vacuum.

Grazing incidence X-ray scattering (GIXS) was performed at the INS instrument installed at the BM32 beam-line of the European Synchrotron Radiation Facility (Grenoble, France), in a ultra-high vacuum system coupled to a Z-axis diffractometer and with a base pressure below 10$^{-10}$~mbar. Several experiments were performed with a monochromatic photon beam of either 11 and 21~keV incident under angles of 0.38 and 0.19$^\circ$ respectively. The corresponding X-ray attenuation length for Ir is 57\,\AA\,at 11~keV, and 68\,\AA\,at 21~keV. The incident beam was doubly focused to a size of 0.4$\times$0.3~mm$^2$ (full width at half-maximum in horizontal and vertical directions, respectively) at the sample location. Detector slits, located 640~mm away from the sample, were set at 1~mm (or 0.5~mm) parallel to the sample surface (which was vertical) and 8~mm perpendicular to it, resulting in angular acceptances of 0.09$^\circ$ (or 0.045$^\circ$) in the vertical direction and 0.7$^\circ$ in the horizontal direction. GIXS data were measured with the help of a standard NaI scintillation point detector.

When used, the reciprocal lattice units (r.l.u.) $h$, $k$, and $l$ are calculated using a hexagonal surface unit cell with lattice vectors $\mathbf{a}_\mathrm{1,s}=1/2 (\mathbf{a}_\mathrm{1,v}+\mathbf{a}_\mathrm{2,v})$, $\mathbf{a}_\mathrm{2,s} = 1/2 (\mathbf{a}_\mathrm{1,v}-\mathbf{a}_\mathrm{2,v})$, $\mathbf{a}_\mathrm{3,s}= \mathbf{a}_\mathrm{1,v} + \mathbf{a}_\mathrm{2,v} + \mathbf{a}_\mathrm{3,v}$, where $\mathbf{a}_\mathrm{1,v}$, $\mathbf{a}_\mathrm{2,v}$ $\mathbf{a}_\mathrm{3,v}$ are the volume unit cell lattice vectors of Ir, all having 3.839\,\AA\, length,  so that surface lattice vectors have 2.714\,\AA\,length in-plane and 6.649\,\AA\, out-of-plane. All scans of the scattered intensity which are plotted and discussed hereafter have been acquired at an out-of-plane reciprocal space lattice unit $l$ = 0.09.

The detector slits (opened 1~mm, located at 640~nm from the sample), intercept a $2\pi/\lambda\times 1/640$ region in reciprocal space along the slits direction, $\lambda$ being the wavelength of the X-ray beam. They are inclined by an angle which equals the scattering angle with respect to the radial scan direction ($\omega$). Due to the slits, the peaks are broadened in radial and azimuthal directions. This broadening varies as a function of the peak order and is systematically corrected for in the following.

\section{Average lattice parameter in graphene}

\begin{figure*}[htpb]
  \begin{center}
  \includegraphics[width=110mm]{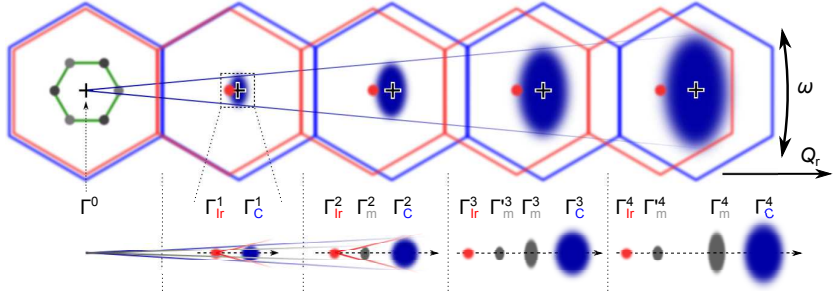}
  \caption{\label{fig:RR}(Color online) Sketch of the first to fifth graphene BZs and their centers for Ir ($\Gamma^\mathrm{0\rightarrow 4}_\mathrm{Ir}$) (red) and graphene ($\Gamma^\mathrm{0\rightarrow 4}_\mathrm{C}$) (blue) where the scattered intensity was measured. Blue ellipses represent in-plane cuts of graphene diffraction rods, characterized by their position, yielding the average projected lattice parameter, as well as by their $Q_\mathrm{r}$ radial and $\omega$ azimuthal widths. For the sake of clarity, the moir\'e diffraction peaks have been omitted from this sketch, but are shown in the close-up view, which marks the measurable moir\'e peaks and gives a geometric construction explaining the observed width of these peaks.}
  \end{center}
\end{figure*}

The in-plane structure of graphene was investigated by exploring an in-plane cut in reciprocal space (Fig.~\ref{fig:RR}), \textit{i.e.} measuring the scattering intensity as a function of the radial $Q_\mathrm{r}$ component of the scattering vector and of the azimuthal angle $\omega$. In order to test the structural quality of graphene, special attention was paid to the moir\'e resulting from the difference in lattice parameter between graphene and Ir(111), which is known to amplify tiny deformations and rotations,\cite{Coraux2008} and is associated to diffraction peaks as we will see.

\begin{figure*}[hbt]
  \begin{center}
  \includegraphics[width=160mm]{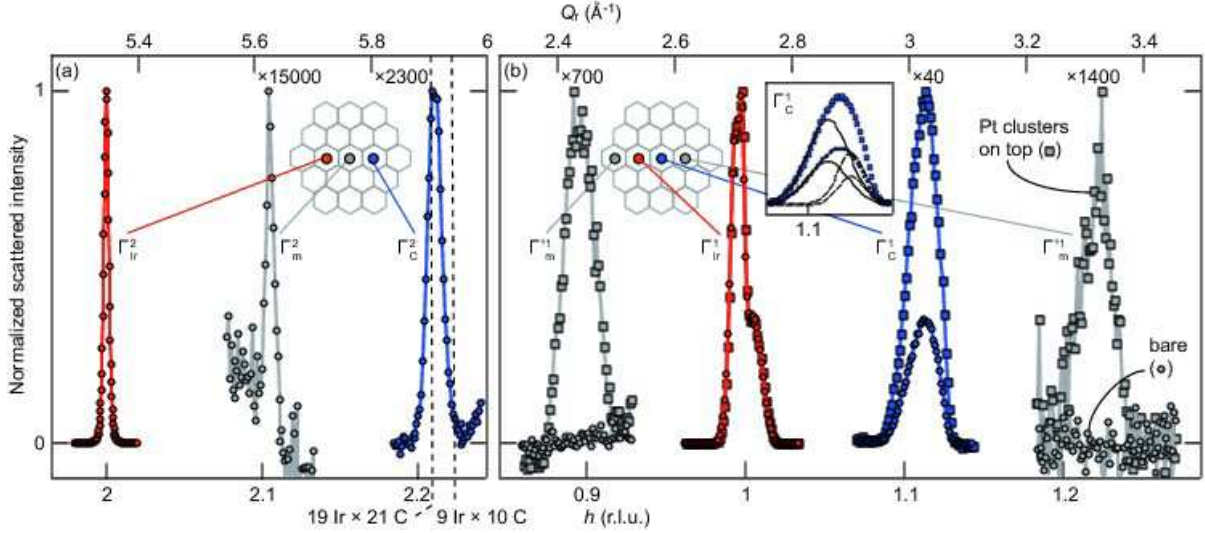}
  \caption{\label{fig:IBZs}(Color online) (a) In-plane radial scans at RT close to the third BZ center for Ir ($\Gamma^\mathrm{2}_\mathrm{Ir}$) and graphene ($\Gamma^\mathrm{2}_\mathrm{C}$) revealing the Ir CTR (red), the graphene rod (blue) and a moir\'e peak ($\Gamma^\mathrm{2}_\mathrm{m}$, gray), as a function of radial momentum transfer ($Q_\mathrm{r}$) and reciprocal space lattice units ($h$). The position of the commensurate structures with 9 Ir cells matching 10 C ones and 19 Ir cells matching 21 C ones are marked. (b) In-plane radial scans measured for another sample (thus having different structure, see section~\ref{sec:T}), with and without Pt clusters on top (square and disk symbols, respectively), close to the second order BZ center for Ir ($\Gamma^\mathrm{1}_\mathrm{Ir}$) and graphene ($\Gamma^\mathrm{1}_\mathrm{C}$), revealing that moir\'e peaks ($\Gamma^\mathrm{,1}_\mathrm{m}$) appear and that the graphene peak is reinforced due to the presence of Pt clusters. Peaks in (b) include two components originating from two Ir crystallites. $\Gamma^\mathrm{1}_\mathrm{C}$ is thus fit with two gaussians, whose centers do not change upon Pt cluster deposition. Insets: sketch of the moir\'e reciprocal lattice.}
  \end{center}
\end{figure*}

At locations where Ir crystal truncation rods (CTRs) or graphene diffraction rods intersect the in-plane cut of the reciprocal space, intensity maxima are found. For the Ir CTR passing through the center of the second ($\Gamma^\mathrm{1}_\mathrm{Ir}$) and third ($\Gamma^\mathrm{2}_\mathrm{Ir}$) Brillouin zones (BZs), the result is shown in Fig.~\ref{fig:IBZs}. The Ir(111) single crystal yields sharp contributions reflecting its high quality. Besides the Ir peaks, a graphene contribution is also observed (at $\Gamma^\mathrm{1}_\mathrm{C}$ and $\Gamma^\mathrm{2}_\mathrm{C}$), whose $Q_\mathrm{r}$ position yields the RT in-plane projection of the lattice parameter of graphene, according to $a_\mathrm{\parallel}$ = $4\pi / (\sqrt{3} Q_\mathrm{r})$, 2.4543$\pm$0.0005\,\AA.

A peak, labeled $\Gamma^\mathrm{2}_\mathrm{m}$ in Fig.~\ref{fig:IBZs}a and referred to as a moir\'e peak in the following, is found halfway between the Ir ($\Gamma^\mathrm{2}_\mathrm{Ir}$) and graphene ($\Gamma^\mathrm{2}_\mathrm{C}$) peaks, corresponding to a $a_\mathrm{m}$ = 25.6$\pm$0.2\,\AA\, period. The origin of this peak is discussed in section~\ref{sec:discussion}. Higher order moir\'e peaks, which are revealed by electron diffraction\cite{Langer2011} thanks to the strong interaction of electrons with matter giving extreme surface sensitivity and multiple diffraction, are not measurable here around $\Gamma^\mathrm{2}_\mathrm{Ir}$; however, as sketched in Fig.~\ref{fig:RR}, other first order moir\'e peaks were measured around $\Gamma^\mathrm{3}_\mathrm{Ir,C}$ and $\Gamma^\mathrm{4}_\mathrm{Ir,C}$.

Platinum deposition on graphene/Ir(111), which yields nanoclusters self-organized on the moir\'e,\cite{NDiaye2009b} increases the intensity of the moir\'e peaks and makes new ones appear (Fig.~\ref{fig:IBZs}b). We note that in the second BZ, the $\Gamma^\mathrm{1}_\mathrm{C}$ position is also the position of the first order moir\'e peak around $\Gamma^\mathrm{1}_\mathrm{Ir}$. This is presumably the origin of the increased intensity at this location in the presence of Pt clusters. The presence of Pt does not alter the position of the graphene peaks (Fig.~\ref{fig:IBZs}b).

\section{Distribution of lattice parameter and crystallographic orientation in graphene}

\begin{figure*}[hbt]
  \begin{center}
  \includegraphics[width=90mm]{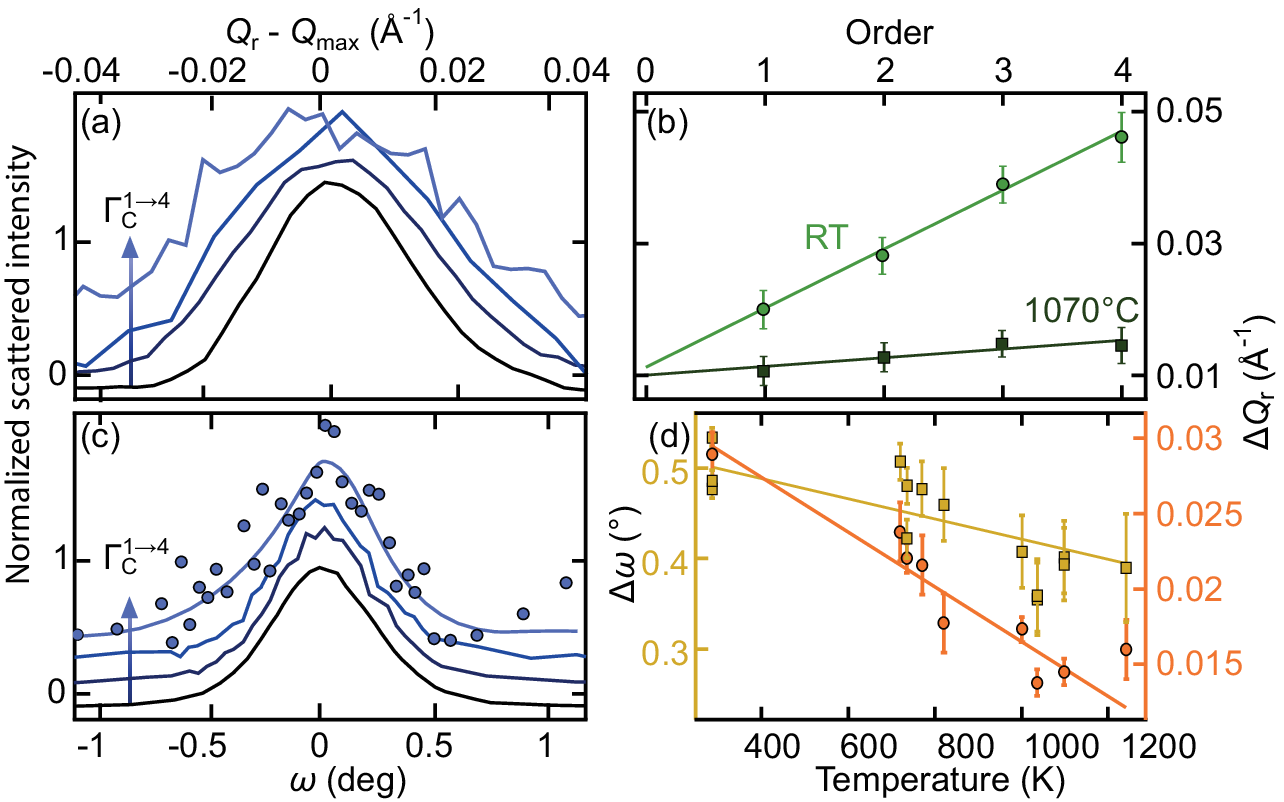}
  \caption{\label{fig:FWHM}(Color online) (a) Radial scans for graphene at different peak orders ($\Gamma^\mathrm{1\rightarrow 4}_\mathrm{C}$), vertically shifted for clarity. (b) Graphene radial scans FWHMs ($\Delta Q_\mathrm{r}$) as a function of peak order ($\Gamma^\mathrm{0\rightarrow 4}_\mathrm{C}$), at RT and 1070~K, and linear fits (lines). (c) Azimuthal angle scans at different peak orders ($\Gamma^\mathrm{1\rightarrow 4}_\mathrm{C}$), vertically shifted for clarity. For $\Gamma^\mathrm{4}_\mathrm{C}$, the azimuthal angle scan (blue disks) has been smoothened (solid lighter blue line) for easing the assessment of the scan FWHM. (d) Graphene radial ($\Delta Q_\mathrm{r}$) and azimuthal ($\Delta \omega$) FWHMs as a function of the sample temperature for the second order peaks.}
  \end{center}
\end{figure*}

The line-shape of diffraction peaks is sensitive to crystal imperfections:\cite{Guinier1994} finite size effects, mosaic spread (in the plane of the sample surface considering the diffraction geometry which we employ), and distributions of lattice parameters all yield peak broadening, the origin of which can be discriminated by studying the broadening dependence on the diffraction peak order. Along the radial direction, the graphene peak width is found to increase linearly with peak order from $\Gamma^\mathrm{1}_\mathrm{C}$ to $\Gamma^\mathrm{4}_\mathrm{C}$ (Fig.~\ref{fig:FWHM}a), in contrast with the $\Gamma^\mathrm{0\rightarrow 4}_\mathrm{Ir}$ Ir peaks, whose radial width is constant (0.01\,\AA$^{-1}$) with order. This implies that the lattice parameter in graphene is spatially varying, with a full width at half maximum (FWHM), given by the increase of the peak width with of peak order (Fig.~\ref{fig:FWHM}b), 0.0093$\pm$0.003\,\AA. The FWHM also has a constant component, obtained by linear extrapolation to zero order (center of the reciprocal space), 0.011\,\AA$^{-1}$, which corresponds to a finite size effect pointing to structurally coherent graphene domains of 60$\pm$30~nm extension.

While their radial width increases as a function of peak order, the graphene peaks have constant azimuthal width (0.53$^\circ$, substantially larger than the constant 0.08$^\circ$ azimuthal width of Ir peaks), as sketched in Fig.~\ref{fig:RR} and shown in Fig.~\ref{fig:FWHM}c, which indicates that small-angle disoriented and/or sheared domains are also present. Similar rotations were reported recently in graphene on Ru(0001), using low energy electron microscopy and micro-LEED.\cite{Man2011} Such rotations should translate into a ca. 10 times larger rotation of the moir\'e peak\cite{Coraux2008} with respect to the nearest Ir peak, \textit{i.e.} into an angular width of the moir\'e peak with respect to the origin equal to that of the graphene peaks divided by its order (Fig.~\ref{fig:RR}a). The measured 0.26$^\circ$ FWHM of the $\Gamma^\mathrm{2}_\mathrm{m}$ moir\'e peak confirms this interpretation. Note that the coexistence of in-plane rotated phases was already reported in the case of large-angle orientation variants in graphene/Ir,\cite{Loginova2009} and small-angle scatter between graphene islands on Ir.\cite{NDiaye2008}

\begin{figure}[hbt]
  \begin{center}
  \includegraphics[width=80mm]{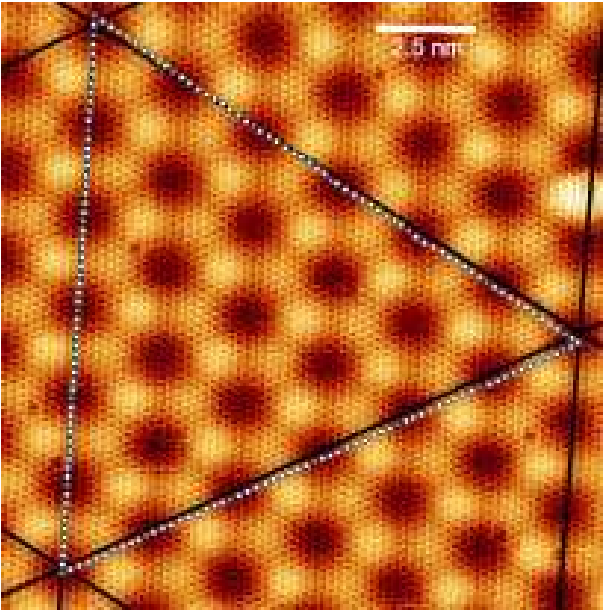}
  \caption{\label{fig:STM}STM topograph of graphene/Ir(111) measured with a 10.9~nA and a 0.2~V tunneling current and bias respectively. Solid lines follow the moir\'e lattice, white dots mark centers of carbon rings. The two triangles share only one edge, evidencing an in-plane shear.
}
  \end{center}
\end{figure}

In the present study, an attempt to further detect lattice variations or distortions by STM could only reveal shears. Shears were detected in STM by scrutinizing the alignment and misalignment on carbon atomic rows and of moir\'e high symmetry directions. Figure~\ref{fig:STM} highlights a triangular region of graphene/Ir(111), into which the carbon zigzag rows align to the high symmetry directions of the moir\'e, only on one of the sides of the regions. The misalignment on the two other sides is of the order of a degree. It is typically one order of magnitude smaller between the graphene and Ir(111) atomic rows. This small shear is in the range estimated by measuring the azimuthal width of the diffraction peaks of graphene. As we will explain in section~\ref{sec:discussion}, detecting the distribution of lattice parameter in graphene by STM is an extremely demanding task, beyond the scope of this article.

\section{\label{sec:T}Dependence of graphene structure on temperature}

\subsection{Dependence on growth temperature}

\begin{figure*}[htpb]
  \begin{center}
  \includegraphics[width=120mm]{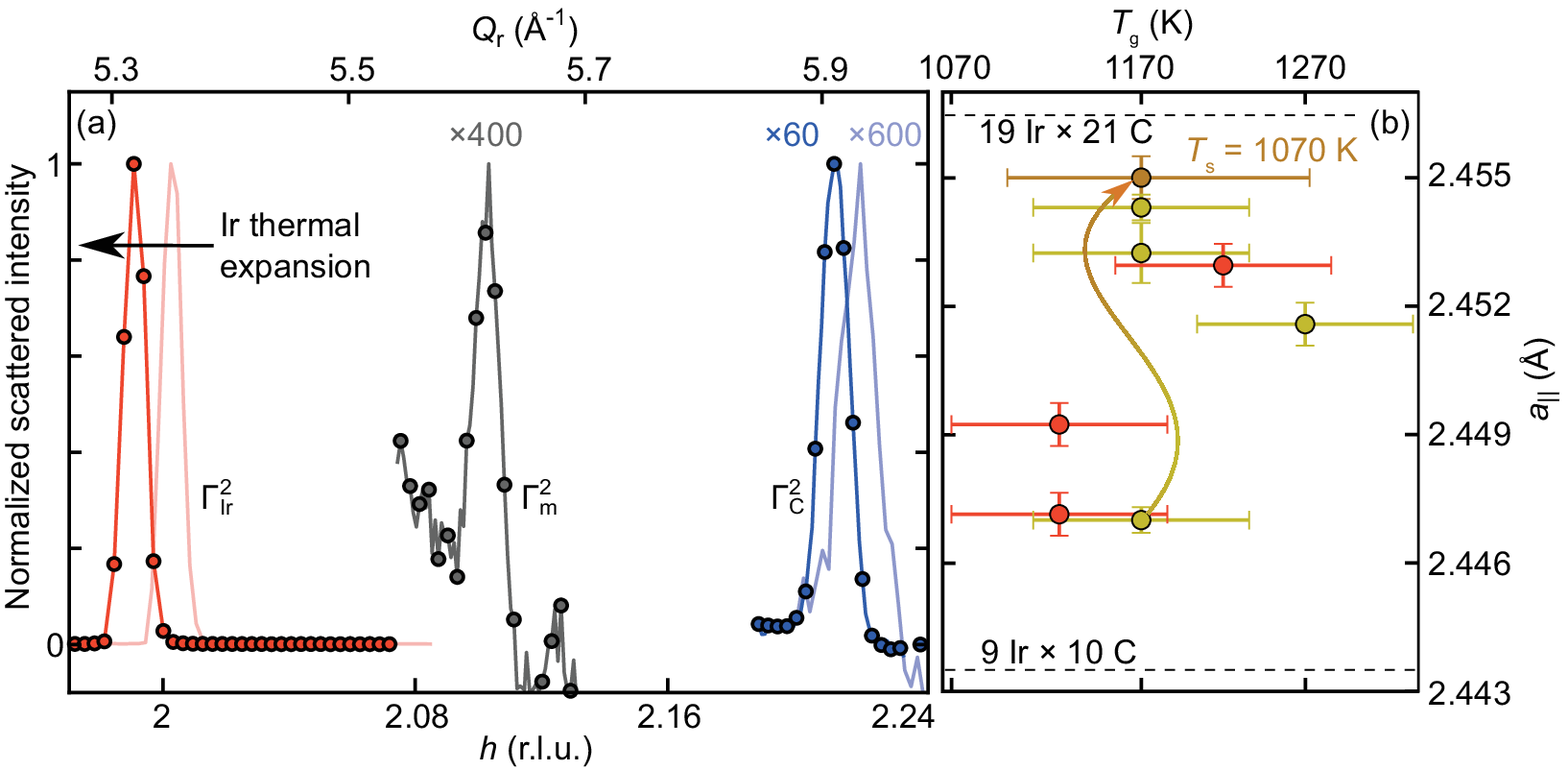}
  \caption{\label{fig:TG}(a) Radial scans close to the third BZ center at 1070~K (light red and light blue), and the same sample at RT (red and blue). Note that the RT measurement corresponds to a sample decorated with Pt clusters reducing counting times (see Fig.~\ref{fig:IBZs}b). (c) Spread of in-plane projections, $a_\mathrm{\parallel}$, of $a_\mathrm{C}$, measured at RT, as a function of the growth temperature $T_\mathrm{g}$, for two series of measurements (yellow and red data points). The orange point corresponds to a sample grown at 1170~K and studied at 1070~K.
}
  \end{center}
\end{figure*}

The average epitaxial matching of graphene on Ir(111) was found to depend sensitively on $T_\mathrm{g}$. Figure~\ref{fig:TG}a shows a graphene peak (light blue) whose position corresponds to $a_\mathrm{\parallel}$ = 2.4470$\pm$0.0005\,\AA, 0.3\% smaller than the previous value, obtained with another sample (see Fig.~\ref{fig:IBZs}a). The accuracy of our temperature probe ($\pm$60~K) does not allow detecting difference in $T_\mathrm{g}$ between the two samples. Actually, the RT graphene lattice parameter $a_\mathrm{\parallel}$ was found to vary by almost 0.01\,\AA\,for different preparations, with no clear dependence on $T_\mathrm{g}$ (Fig.~\ref{fig:TG}b).

Discrepancies were found also in the value of the radial width of the graphene peaks, between two samples tentatively prepared at the same temperature, within the accuracy of the temperature probe. These variations are accounted for in Fig.~\ref{fig:FWHM}d in the form of error bars. The general trends, namely the FWHM slope and extrapolation to the center of reciprocal space, do not vary significantly. This implies that the two samples have similar domain size and width of the distribution of graphene lattice parameters. We found no clear dependence of the graphene radial width neither on $T_\mathrm{g}$ nor on $a_\mathrm{\parallel}$, in line with the behavior of $a_\mathrm{\parallel}$ \textit{versus} $T_\mathrm{g}$.

\subsection{Dependence on sample temperature}

Varying $T_\mathrm{s}$ away from $T_\mathrm{g}$ also influences the epitaxial matching: while between RT and 1070$\pm$60~K the Ir lattice parameter is expanded by 0.48\%, for graphene $a_\mathrm{\parallel}$ is increased by only 0.33\%, from 2.4470$\pm$0.0005 to 2.455$\pm$0.001\,\AA\,(Figs.~\ref{fig:TG}a,b). The superperiodicity is slightly decreased, to 24.5$\pm$0.2\,\AA. The radial scans widths (Fig.~\ref{fig:FWHM}b) are again found to increase with peak order, but to a much lesser extent than at RT. The corresponding distribution of lattice parameter, 0.0023\,\AA, is almost one order of magnitude smaller than at RT, but the extrapolation to zero of the radial scans width yields a value for the size of the structurally coherent graphene domains similar to the RT value, 60$\pm$30~nm. The angular widths are also constant (0.43$^\circ$) as a function of order, but smaller than at RT. Figure~\ref{fig:FWHM}d shows that both the radial and angular widths of a graphene peak increase with decreasing the sample temperature.

\section{\label{sec:discussion}Discussion}

\begin{figure}[htpb]
  \begin{center}
  \includegraphics[width=66mm]{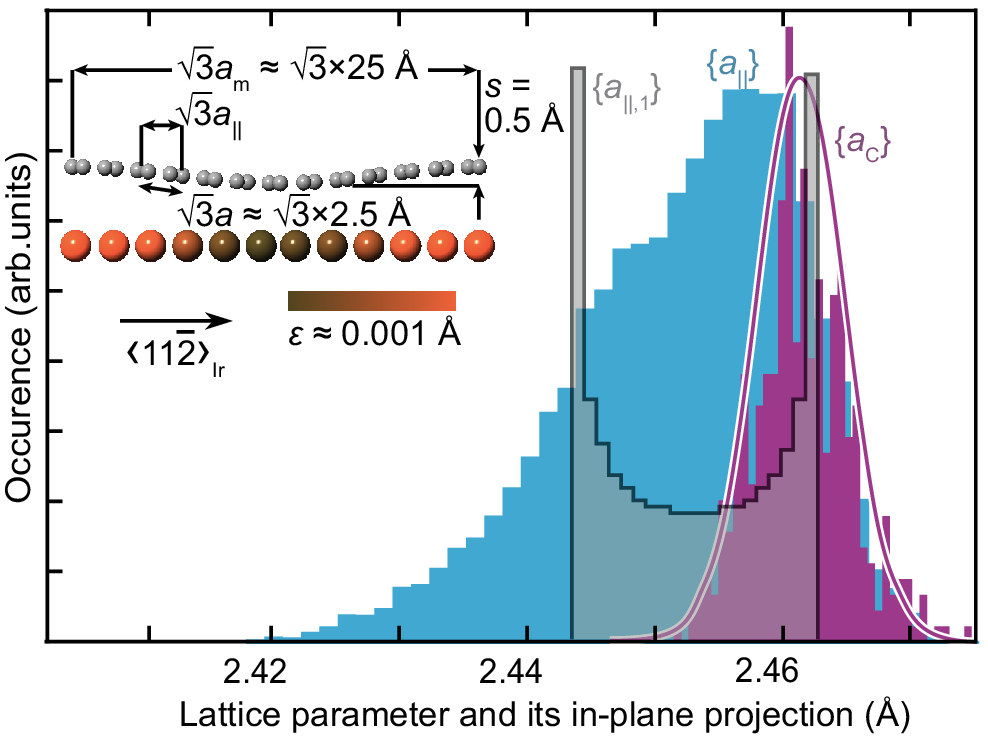}
  \caption{\label{fig:nanorippling}(Color online) Construction of the distribution of projections of lattice parameters in graphene, $\left\lbrace a_\mathrm{\parallel}\right\rbrace$, from the distribution of lattice parameters,  $\left\lbrace a_\mathrm{C}\right\rbrace$, taking the nanorippling of graphene along the moir\'e into account. The purple histogram shows a distribution of graphene lattice parameters randomly chosen in a standard normal (Gaussian) distribution (as a guide for the eye, the corresponding gaussian function is plotted with a purple line), with FWHM corresponding to the increase of the width of the graphene peak as a function of BZ order in the radial direction (Fig.~\ref{fig:FWHM}b). The grey histogram is the distribution of in-plane projections yielded by nanorippling for one specific value of the lattice parameter. The cyan histogram includes all such distributions each corresponding to all lattice parameter values included in the distribution shown in the purple histogram. The distributions were derived by taking into account 10 moir\'e unit cells in a one-dimensional model. The inset shows a side-view of this model along a $\left\langle11\bar{2}\right\rangle$ direction in Ir illustrating the nanorippling-induced $\left\lbrace a_\mathrm{\parallel,1}\right\rbrace$ distribution for a specific $a_\mathrm{C}$, and the periodic strain in the last Ir plane (red shades, $\varepsilon$ amplitude). The characteristic lengths of the model, the moir\'e lattice parameter $a_\mathrm{m}$, the graphene lattice parameter $a$, its in-plane projection $a_\mathrm{\parallel}$, and the nanorippling amplitude $s$ are shown. The $\sqrt{3}$ factor corresponds to the $\left\langle11\bar{2}\right\rangle$ direction.}
  \end{center}
\end{figure}

The RT lattice parameter of isolated graphene --a conceptual object-- was calculated to be $a_\mathrm{0,RT}$ = 2.4565\,\AA.\cite{Zakharchenko2009} In the above we have extracted the in-plane projection of the lattice parameter in graphene, $a_\mathrm{\parallel}$. Deducing from this parameter the actual lattice parameter, $a_\mathrm{RT}$, requires taking the effect of the graphene/Ir moir\'e into account. The weak, varying interaction between graphene and Ir as a function of the location in the moir\'e (see Fig.~\ref{fig:nanorippling}) is indeed associated with a nanorippling of graphene, characterized with a $s$ =0.5\,\AA\ typical height amplitude.\cite{Busse2011} Assuming a sinusoidal variation of the height, $a_\mathrm{RT}$ is typically 0.3\% larger than $a_\mathrm{\parallel}$. This yields $a_\mathrm{C}$ = 2.4617\,\AA\, and 2.4548\,\AA\, for the samples studied in Figs.~\ref{fig:IBZs}a and \ref{fig:TG}a respectively, \textit{i.e.} 0.2\% larger and 0.1\% lower than $a_\mathrm{0,RT}$ respectively. The self-organization of Pt clusters on top of graphene does not modify the in-plane component of the lattice parameter of graphene, however it could affect the nanorippling, in turn changing its lattice parameter. The small deviations of  $a_\mathrm{C}$ from $a_\mathrm{0,RT}$ are consistent with the weak interaction between graphene and Ir(111), characterized by a limited charge transfer between graphene and Ir, typically 0.01 electron or hole per atom.\cite{Busse2011}

By contrast, in graphene/Ru(0001), a much larger $a_\mathrm{\parallel}$ = 2.4895\,\AA\,value was found.\cite{Martoccia2008} In this system, nanorippling is more pronounced than in graphene/Ir(111), corresponding to an expected height variation of 1.5\,\AA\,\cite{Wang2008}, which yields $a_\mathrm{C}$ roughly 2\% larger than $a_\mathrm{0,RT}$, consistent with the strong electron doping\cite{Wang2008,Sutter2008} of graphene on Ru.

The observed weak moir\'e peaks are associated to the weak but non vanishing graphene-Ir(111) interaction. We assume a small in-plane sinusoidal strain in Ir (sketched in Fig.~\ref{fig:nanorippling} with a color gradient for Ir atoms) stemming from this interaction, having locally a weak tendency to covalent bonding,\cite{Busse2011} which according to density functional theory calculations moves Ir atoms away from their position in bare Ir(111) by a $\varepsilon$ value of a few 0.001\,\AA\, with a  $a_\mathrm{m}$ period\cite{privateBusse} as a function of their position in the moir\'e lattice. In a simple one-dimensional model, from the position of an Ir atom labeled $i$ in a chain, $a_\mathrm{Ir}\times (i+\varepsilon\cos(2\pi i a_\mathrm{Ir}/a_\mathrm{m}))$, the scattered amplitude calculated by summing up the contribution of all Ir atoms accounts for the observed so-called moir\'e peaks, whatever the relation, commensurate or not between the Ir, graphene and moir\'e lattice parameters. The periodic modulation of $a_\mathrm{C}$ in graphene also contributes to the moir\'e peak intensities, but to a much smaller extent, since scattering by C atoms is ca. 40 times lower than by Ir ones.

The measured $a_\mathrm{\parallel}$ values are substantially larger than the value (2.4435\,\AA) expected for a commensurate phase with a moir\'e consisting of 10 graphene cells matching 9 Ir ones. Second order commensurability, with 25 graphene unit cells matching 23 metal ones, was concluded for graphene/Ru(0001).\cite{Martoccia2008} The closest high order (second and third) graphene/Ir(111) commensurate moir\'e would consist of 21 graphene unit cells matching 19 metal ones, which is yet slightly but significantly off the $a_\mathrm{m}$ and $a_\mathrm{\parallel}$ values. We conclude that on the average, at RT, the graphene and Ir lattices are incommensurate.

However, a description in terms of a single incommensurate graphene/Ir(111) phase would be too naive as a distribution of lattice parameters was unveiled. This distribution is plotted in Fig.~\ref{fig:nanorippling}. It is tempting to invoke the periodic modulation of $a_\mathrm{\parallel}$ due to the moir\'e nanorippling (see Fig.~\ref{fig:nanorippling}) as the interpretation of this distribution. Nevertheless, these variations are periodic (or quasi-periodic in case of incommensurate structure) and thus contribute to all observed peaks without yielding any broadening of the graphene peaks. Small random or uncorrelated displacements similar to thermal motion, either static or dynamic, are also excluded as they only yield a Debye-Waller like decrease of the peak height with order, without peak widening.\cite{Guinier1994} Only two scenarios are actually relevant for accounting for the distribution of lattice parameter: the existence of domains with different average lattice parameters (first scenario) or progressive changes of lattice parameter within large domains (second scenario). These scenarios are discussed more into details at the end of this section in the light of the effects of temperature, which we will now address.

The fact that the RT lattice parameter $a_\mathrm{C}$ varies by several tenth of a percent as a function of the growth temperature $T_\mathrm{g}$, without simple relationship with it, suggests that due to the mismatch in thermal expansion coefficient (TEC) between graphene and Ir,\cite{NDiaye2009} the epitaxy of graphene on Ir(111) is initially set at $T_\mathrm{g}$, and that the strain induced during cooling down to RT by the mismatch in TEC might modifies the epitaxy in a non trivial way.

Taking nanorippling into account as before, we find that at 1070$\pm$60~K $a_\mathrm{C}$ is 0.3\% larger than the expected lattice parameter ($a_\mathrm{0,1070}$ = 2.4556\,\AA) for isolated graphene at this temperature.\cite{Zakharchenko2009} At RT for this sample, the deviation to the expected $a_\mathrm{RT}$ value was less than 0.1\%. The average epitaxial relationship thus indeed changed between 1070~K and RT, which suggests some slippage of graphene on Ir(111) against temperature-induced stress, presumably due to the high stiffness of graphene. The TEC of graphene is found positive, at variance with the vanishing\cite{Zakharchenko2009} or negative\cite{Mounier2005} lattice parameter variation predicted for free-standing graphene in this temperature range. The TEC is much larger than estimated from first principle calculations for graphene/Ir(111).\cite{Pozzo2011} The partial inheritance, in graphene, of the Ir TEC, could result from the weak tendency to periodic chemical bonding in graphene/Ir (Ref.~\onlinecite{Busse2011}), or from more subtle effects near steps edges (see below). A more extreme situation was encountered in a related system, $h$-BN on Rh(111) (Ref.~\onlinecite{Martoccia2010}) in which the two TECs were found equal presumably due to the very strong interaction between the two materials.

Graphene not only exhibits domains with different structures (see discussion below) at RT, but also at 1070~K (Fig.~\ref{fig:FWHM}b). The fact that at this temperature wrinkles are absent\cite{NDiaye2009} suggests that the domains develop during growth. The spread in the structure between domains is however amplified as the temperature decreases from $T_\mathrm{g}$ (Fig.~\ref{fig:FWHM}d). This suggests that graphene minimizes its elastic energy by amplifying the structural variations as the compressive epitaxial stress, due to a smaller TEC than that of Ir, increases.

This brings us to the discussion about the two above-mentioned possible scenarios.

\begin{figure}[hbt]
  \begin{center}
  \includegraphics[width=80mm]{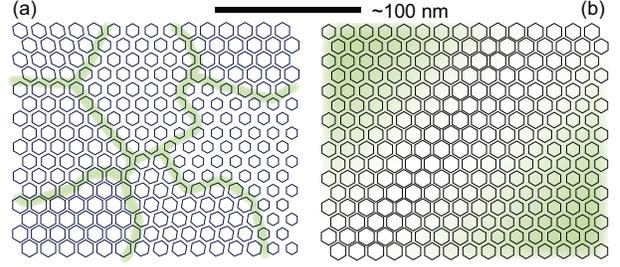}
  \caption{\label{fig:scenarios}(Color online) Graphene deformations mapped onto a regular mesh with hexagons at its nodes, whose shapes represent amplified deformations (strain, rotation, shear), for two cases: ca. 60~nm domains, each having distinct lattice parameter (a), and a domain wall (ca. 100~nm) between two domains having the same lattice parameter (b).}
  \end{center}
\end{figure}

The first scenario is sketched in Fig.~\ref{fig:scenarios}a. It shows domains with a size of ca. 60~nm, most being incommensurate with Ir(111), possibly coexisting with commensurate phases (21 C hexagons coinciding with 19 Ir distances for sample of Figs.~\ref{fig:IBZs}a and 11 C hexagons matching 10 Ir for sample in Fig.~\ref{fig:TG}a) present in small proportion. The coexistence of various graphene phases in this scenario implies the existence of a large number of shallow minima in the energy of graphene upon nanorippling, bond streching/compression, shearing, rotations, due to a complex balance between graphene's flexural modulus and interaction with Ir.

The second scenario is sketched in Fig.~\ref{fig:scenarios}b, which shows several 10~nm domains, commensurate, thus pinned to the Ir(111) lattice, all having the same structure, but separated by domain walls, so-called discommensurations.\cite{Guinier1994} The strain field in these walls fixes their size: the 0.0093\,\AA\,lattice parameter distribution FWHM suggests that one (or 1/3 or 2/3 if the domains are located on different terraces) Ir(111) lattice parameter is accommodated in a few 100 graphene unit cells, \textit{i.e.} typically 100~nm. We also note that in this second scenario, the typical size of the graphene domains or domain walls is of the order of the average distance of ca. 100 nm between Ir step edges (Fig.~\ref{fig:AFM}). This may not be fortuitous: Ir step edges, where the interaction between graphene and Ir is expected to be the strongest, could pin the graphene lattice, and drive the formation of commensurate graphene domains. Beyond a critical size, growing these commensurate domains would cost too high elastic energy with regard to the low interaction with Ir, so that inhomogeneous deformation and shears/rotations resulting in incommensurability between steps are preferred. This would result in large discommensurations extending over the terraces. The increase of inhomogeneous strain and rotations with decreasing temperature would result naturally from the increasing lattice parameter mismatch with respect to the growth temperature. This scenario would more easily explain than the first one the partial inheritance of the Ir TEC by graphene, as this could result from a complete inheritance in the small commensurate domains near step edges.

Detecting the variations of lattice parameters discussed above with STM is an extremely demanding task, and our attempts in this direction remain unsuccessful. In the fist scenario, this would require detecting variations of $a_\mathrm{C}$ of a few 0.001\,\AA\, (the FWHM of the distribution of $a_\mathrm{C}$ is 0.0093\,\AA, see above), or of a few 0.01\,\AA\, taking benefit of the ca. 10-fold amplification of strains by the moir\'e. This is already below or at the verge of the resolution of the STM instrument which has been employed whose resolution is limited to 0.004\,\AA\, and 0.04\,\AA\, precisions for direct $a_\mathrm{C}$ measurements and moir\'e analysis respectively,\cite{NDiaye2008} which are typical values for high resolution STM. In addition, such variations should be detected between graphene domains distant by several 10~nm (first scenario), or continuously over ca. 100~nm-wide discommensurations (second scenario), which would require drift effects to be less than a few 0.001\,\AA\, and 0.01\,\AA\, for direct $a_\mathrm{C}$ analysis and moir\'e analysis respectively, another requirement which cannot be reached in most STM instruments. We note that micro-diffraction, such as possible in a low-energy electron microscope, has been used to detect ca. 0.002\,\AA\, variations of $a_\mathrm{C}$ in graphene/Ru(0001).\cite{Man2011} This technique does not however provide sufficient lateral resolution (set by the diameter of the aperture employed for defining the size of the micro-spot) for detecting the variations in graphene/Ir(111), which are characterized by length-scales in the range of several 10~nm.

\section{Conclusion}

The carbon bond length in graphene on Ir(111) is equal to that calculated for isolated graphene to within 0.2\%. High resolution X-ray scattering measurements reveal that the structural parameters in very high quality graphene have some scatter, 0.3\% regarding lattice parameter and 0.5$^\circ$ regarding in-plane rotations. These small variations are either due to the presence of domains, with size ca. 60~nm and different structure, most often incommensurate with Ir(111), or to discommensurations, extending across ca. 100~nm, between graphene domains pinned to the substrate defects and commensurate with it. We anticipate that such structural variations may also be encountered in other systems weakly interacting with their substrate, for instance graphene on Cu or dichalcogenates prepared by chemical vapor deposition. The structure of graphene, average lattice parameter, rotations and strain, is found to depend sensitively on temperature, which could allow for epitaxial control of its properties. Understanding the influence of the observed structural variations over particle or quasiparticle scattering, phonon modes, local charge transfers and chemical inhomogeneities, which steer many of graphene's properties, call for further investigations. We anticipate that intercalation of species suppressing the non-vanishing graphene-metal interaction, like alkali metals or hydrogen, could yield truly free-standing graphene where the structural inhomogeneities could be released.

\section{Acknowledgements}

We thank T. Michely for the analysis of Fig.~\ref{fig:STM} and C. Busse for STM measurements and fruitful discussions. C. V.-V. acknowledges support from the Fondation Nanosciences. Research supported by EU NMP3-SL-2010-246073 "GRENADA" and French ANR ANR-2010-BLAN-1019-NMGEM contracts.

%\bibliography{graphene_Ir_strains}


\end{document}